\documentclass[conference]{IEEEtran}
\usepackage{lipsum}
\usepackage{tikz} \usetikzlibrary{positioning} \tikzset{font=\footnotesize}
\usepackage{standalone}
\usepackage{graphicx} % Required for inserting images
\usepackage{svg}

\usepackage{color}
\usepackage{subcaption} % Required for subfigures
\usepackage{makecell}

\usepackage{cite}
\usepackage{amsmath,amssymb,amsfonts}
\usepackage[ruled, lined, linesnumbered, commentsnumbered, longend]{algorithm2e}
\usepackage{algpseudocode}

\usepackage{graphicx}
\usepackage{textcomp}
\usepackage{xcolor}
\usepackage{soul}
\usepackage{enumerate}
\usepackage{comment}

\usepackage{subcaption}
\usepackage{multirow}
\usepackage{tabu}
\usepackage{float}
\usepackage{booktabs}
\usepackage{arydshln}
\usepackage[inline]{enumitem}

\usepackage{amsmath}
\usepackage{mathtools}
\usepackage{textcomp}
\usepackage{xurl}

\definecolor{boristext}{rgb}{0.22, 0.44, 0.33}
\definecolor{boriscomments}{rgb}{0.83, 0.0, 0.0}
\definecolor{davidcomments}{rgb}{0.0, 0.0, 0.83}
\definecolor{miguelcomments}{rgb}{0.5, 0, 0.8}

\begin{document}
\title{Can cloud-based VR streaming handle Wi-Fi OBSS contention?}

\author{Miguel Casasnovas$^{\flat}$, Marc Carrascosa-Zamacois$^{\star}$, Boris Bellalta$^{\flat}$\\
$^{\flat}$ Wireless Networking group. Universitat Pompeu Fabra, Barcelona, Spain.\\
$^{\star}$ Centre Tecnològic de Telecomunicacions de Catalunya, Barcelona, Spain.}

\maketitle

\begin{abstract}
This paper experimentally analyzes the negative impact of contention caused by neighboring Wi-Fi networks operating on overlapping channels on Virtual Reality~(VR) streaming over Wi-Fi, focusing on scenarios of partial and full channel overlap within an 80~MHz channel. Our results show that $(i)$~increasing the number of 80~MHz Overlapping Basic Service Sets (OBSSs) intensifies contention and degrades VR streaming performance; $(ii)$~OBSS activity on the secondary-sided 40~MHz portion degrades performance more than activity on the primary-sided 40~MHz portion;  $(iii)$~for the same aggregate load, full channel overlap with two 40~MHz OBSS contenders is less detrimental than partial overlap with a single high-load 40~MHz contender, but more disruptive than full overlap with two 80~MHz contenders; and $(iv)$~full channel overlap with two 40~MHz OBSS contenders has a smaller impact on VR streaming under symmetric traffic loads than under asymmetric loads. Moreover, our results demonstrate that our previously proposed Network-aware Step-wise adaptive bitrate algorithm for VR streaming~(NeSt-VR) effectively mitigates performance degradation in OBSS environments, enabling VR streaming under heavier OBSS traffic conditions.
\end{abstract}


%------------------------------------
%------------------------------------
%------------------------------------
%------------------------------------

\section{Introduction}
IEEE 802.11-based wireless local area networks~(WLANs), commonly known as Wi-Fi networks, have been increasingly deployed due to their cost-effectiveness, mobility, flexibility, and capacity to satisfy the bandwidth and low-latency requirements of sensitive, bandwidth-hungry applications such as eXtended Reality (XR)---encompassing Virtual, Augmented, and Mixed Reality (VR, AR, MR)~\cite{galati2024will}.

Wi-Fi networks operate in shared, unlicensed frequency bands. Since the number of channels is limited in the radio frequency spectrum, Overlapping Basic Service Set~(OBSS)\footnote{A Basic Service Set (BSS) comprises a single Access Point (AP) and one or more associated clients.}
 scenarios are common in dense Wi-Fi deployments (e.g., residential buildings, enterprise environments, and public areas)~\cite{bellalta2016ieee}. Overlapping coverage areas can lead to increased medium contention and reduced airtime utilization, resulting in performance degradation, such as reduced throughput and increased latency~\cite{dunat2004impact}---issues that are particularly detrimental to delay-sensitive applications requiring consistent and ultra-high bandwidth connections, such as XR streaming.
While OBSS contention has been extensively studied in general Wi-Fi contexts---e.g., Faridi et al.~\cite{faridi2016analysis} modeled OBSS interactions using continuous Markov chains, and Deek et al.~\cite{deek2011impact} showed that partial frequency overlap can degrade performance more than full overlap---the specific impact of OBSSs on VR streaming remains largely unexplored, despite being recognized as a critical research gap~\cite{hossain2023survey}. 
Indeed, to the best of our knowledge,~\cite{mehrnoush2022ar} is the only existing study evaluating OBSS contention in VR contexts; the authors relied on simulations to quantify the impact of OBSSs on the number of supported VR headsets. Thereby, no prior work has experimentally analyzed VR streaming performance under OBSS contention. 

\begin{figure}[t!!!]
    \centering
    \includegraphics[width=.9\linewidth]{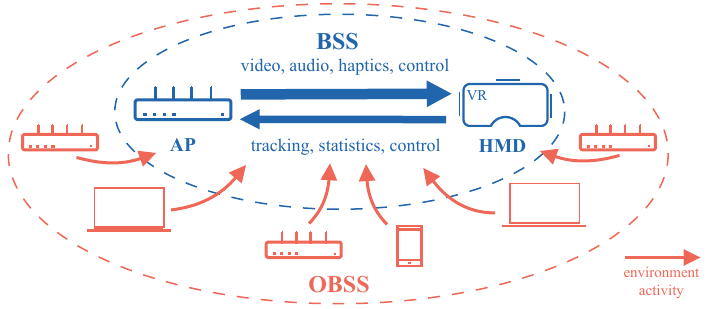}
    \caption{{Considered dense deployment scenario where a VR BSS coexists with multiple interfering OBSSs}.}
    \label{fig:intro_fig}
\end{figure}

This paper addresses this gap by experimentally investigating the impact of inter-BSS transmissions---originating from BSSs operating on fully or partially overlapping frequency bands and thereby generating OBSS contention---on VR streaming over Wi-Fi~6 in the 5~GHz band (see Fig.~\ref{fig:intro_fig}). It further evaluates the effectiveness of Adaptive BitRate (ABR) algorithms in mitigating the resulting performance degradation.
In particular, our evaluation is carried out using Air Light VR~(ALVR)\footnote{\url{https://github.com/alvr-org/ALVR}}, an open-source solution designed for streaming VR content over Wi-Fi, and examines the practical impact of OBSS activity on VR streaming and the effectiveness of the Network-aware Step-wise ABR algorithm for VR streaming (NeSt-VR)~\cite{maura2024experimenting, casasnovas2025nest} in maintaining VR streaming performance under OBSS contention scenarios.

%--------------------------------------------------------------------

The main contributions of this paper are:
\begin{enumerate}
    \item Studying experimentally the impact of OBSS contention on VR streaming performance over Wi-Fi. Considering both partial and full overlapping scenarios.
    \item Demonstrating that effective VR-tailored ABRs, such as NeSt-VR, substantially reduce network delays and frame losses, thereby meeting the stringent latency and reliability requirements of VR streaming in dense WLAN deployments under OBSS contention.
    \item Quantifying how much OBSS traffic is required to disrupt VR streaming traffic, revealing the surprising resilience of VR streaming to handle high OBSS traffic levels, especially when using NeSt-VR.
\end{enumerate}

%------------------------------------
%------------------------------------
%------------------------------------
%------------------------------------
\section{VR streaming over Wi-Fi}\label{sec:background}
%------------------------------------
%------------------------------------
\subsection{VR streaming}\label{sec:bg_vr_streaming}

%------------------------------------
\subsubsection{ALVR}
ALVR is an open-source project that enables wireless VR streaming from a server to a standalone VR headset over Wi-Fi.
The server (streamer) handles game execution and scene rendering, while the headset (client) manages sensor tracking and display.
Tracking data is collected by the client and sent upstream to the server. The server uses this data for pose prediction, runs the game logic, renders the scene, and encodes it into a video frame. This frame is then fragmented into multiple packets and transmitted downstream to the client using either a constant bitrate or an adaptive bitrate approach.
Upon reception, the client reassembles and decodes the video frames, which are then rendered and displayed to the user.
Besides video, the server sends audio, haptic feedback, and control data downstream, while the client transmits tracking, statistics, and control data upstream, as illustrated in Fig.~\ref{fig:intro_fig}.

%------------------------------------
\subsubsection{NeSt-VR}

NeSt-VR, introduced in~\cite{maura2024experimenting} and extended in~\cite{casasnovas2025nest}, is a configurable ABR algorithm designed to optimize VR streaming over Wi-Fi by dynamically adjusting the video bitrate based on real-time network conditions, ensuring low latency and a smooth user experience. This algorithm has been integrated into a custom ALVR fork\footnote{\url{https://github.com/wn-upf/NeSt-VR/tree/extension_changes}} and has demonstrated effectiveness in both single-user and multi-user scenarios, in both lab and real-world environments.
As described in~\cite{casasnovas2025nest}, the algorithm adjusts probabilistically the bitrate in discrete steps to avoid abrupt quality shifts that could degrade the user's experience. Adaptations are based on the frame delivery success rate---measured via the Network Frame Ratio (NFR), which reflects the proportion of successfully delivered frames---and the network delay---measured via the Video Frame Round-Trip Time (VF-RTT), which captures the round-trip latency of video frames.\footnote{Recommended thresholds to ensure a positive user experience in VR streaming---regardless of the targeted frame rate---are an average NFR of 99\% and an average VF-RTT of 33~ms~\cite{casasnovas2025nest}.} It also avoids overreacting to transient congestion by smoothing out short-term fluctuations through measurement averaging and ensures that the accurately estimated available network capacity is not exceeded.
%------------------------------------
%------------------------------------
\subsection{Wi-Fi 6}\label{sec:wifi}

%------------------------------------

Wi-Fi~6 (IEEE 802.11ax)~\cite{khorov2018tutorial} supports channel bandwidths up to 160~MHz by aggregating contiguous 20~MHz channels through \textit{Channel Bonding}. A 20~MHz channel is designated as the \textit{primary} (control) \textit{channel} and the remaining bonded channels are considered \textit{secondary channels}.

To avoid collisions, devices perform \textit{Clear Channel Assessment}~(CCA) before transmitting. In particular, if the primary channel is sensed as busy, the device defers its transmission. 
CCA deferrals are particularly common in dense deployments since multiple BSSs often operate on overlapping frequencies, increasing contention and reducing transmission opportunities. Moreover, \textit{Adjacent Channel Interference}~(ACI), caused by transmissions on neighboring, non-overlapping channels, may increase CCA deferrals, as signal leakage into adjacent frequency bands leads devices to sense the channel as busy. 

Nonetheless, OBSSs can still significantly impair performance---especially for bandwidth- and latency-sensitive applications such as VR streaming---due to persistent inter-network contention, frequent transmission deferrals, and underutilization of available spectrum.

%------------------------------------
%------------------------------------
%------------------------------------
%------------------------------------
\section{Experimental Setup}\label{sec:vr_streaming}

%------------------------------------
%------------------------------------

\subsection{Considered scenarios}\label{sec:scenarios}

\begin{figure}[t]
    \centering
    \includegraphics[width=.9\linewidth]{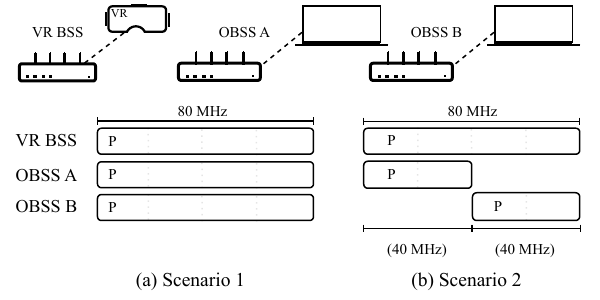}
    \caption{Illustration of the considered OBSS scenarios. %Scenario~1: Coexistence with 80~MHz OBSSs. Scenario~2: Coexistence with 40~MHz OBSSs.
    }
    \label{fig:scenarios}
\end{figure}

Fig.~\ref{fig:scenarios} provides a graphical representation of the two scenarios considered in this work. 
Both scenarios include a BSS dedicated to VR streaming (hereinafter referred to as the VR~BSS), operating on an 80~MHz-wide channel in the 5~GHz band---as recommended in~\cite{michaelides2025lessons}---and using channels 52--64~(with 52 as the primary), and two OBSS contenders (hereinafter referred to as OBSS~A and OBSS~B):
\begin{itemize}
    \item \textit{Coexistence with 80 MHz OBSSs~(Scenario~1)}: OBSSs~A and~B operate on the same 80~MHz channel as the VR~BSS, i.e., channels 52--64~(with 52 as the primary).
    \item \textit{Coexistence with 40 MHz OBSSs~(Scenario~2)}: OBSSs~A and~B operate on adjacent 40~MHz channels that partly overlap with the VR~BSS's 80~MHz channel. OBSS~A uses channels 52--56~(with 52 as the primary), while OBSS~B uses channels 60--64~(with 60 as the primary). Thus, each OBSS causes contention in a distinct half of the VR~BSS's 80~MHz channel: OBSS~A on the primary-sided half and OBSS~B on the secondary-sided half.
\end{itemize}

%------------------------------------
%------------------------------------

\subsection{Test-bed}\label{sec:setup}

Thirty-second tests were conducted in a controlled indoor environment at the Department of Engineering of Universitat Pompeu Fabra (UPF), for both Scenario~1 and Scenario~2. Detailed equipment specifications can be found in Table.~\ref{tab:equipment}.

The VR streaming system comprised a high-performance, VR-ready PC acting as the streaming server, and a Meta Quest~2 Head-Mounted Display (HMD) functioning as the client, both connected to a dedicated gaming-centric AP. 
The SteamVR Home environment was streamed at 90~fps using a modified version of ALVR v20.6.0 (see \cite{maura2024experimenting, casasnovas2025nest}), either at a constant bitrate (CBR) of 100~Mbps or using NeSt-VR’s \textit{Balanced} profile, which dynamically adjusts the bitrate between 10 and 100~Mbps, starting at 100~Mbps.

On the other hand, OBSS~A and OBSS~B APs, both running the OpenWrt\footnote{\url{https://openwrt.org/}} firmware, generated controlled OBSS load. Each OBSS AP delivered downlink UDP\footnote{\texttt{iperf3}'s UDP traffic option enables fixed-rate load control, while its TCP option sends as much data as the network can handle.} traffic using \texttt{iperf3}\footnote{\url{https://iperf.fr/}} from a dedicated laptop server (model a) to a dedicated laptop client (model b). Distinct traffic levels were tested across experimental runs, including asymmetric and symmetric traffic configurations. 

Servers were connected to their APs via 1~Gbps Ethernet, while clients connected wirelessly over Wi-Fi~6 in the 5~GHz band, with no multi-user features (such as OFDMA or MU-MIMO) enabled, as recommended in~\cite{michaelides2023wi,michaelides2025lessons}. Each wireless link used two spatial streams and up to Modulation and Coding Scheme (MCS) 11, enabling high-throughput. 
In particular, Received Signal Strength Indicator (RSSI) values remained below -45~dBm, with a noise floor near -92~dBm. 

APs and stations were positioned in close proximity, minimizing classic problems such as the hidden node problem (where a device cannot sense another ongoing transmission, leading to collisions) and the exposed node problem (where a device unnecessarily defers transmission due to a nearby non-interfering signal). Nevertheless, ACI may still pose a significant source of contention and interference in Scenario~2.

\begin{table}[t]
    \centering
    \caption{Equipment details.}
    \footnotesize
\begin{tabular}{@{}lll@{}}
\toprule
     \textbf{1x PC (vr)} & OS       & Windows 10 x64 \\
    & GPU  & NVIDIA GeForce RTX 3080, 10 GB  \\
    & CPU                      &  i5-12600KF  \\
    \midrule
    \textbf{2x HMD} & Model & Meta Quest 2 \\
    \midrule
    \textbf{2x laptop a} & Model & Dell Latitude 3520 \\
    & OS       & Ubuntu 22.04.5 LTS \\
    & GPU  & Iris Xe Graphics  \\
    & CPU  &   i7-1165G7  \\
    \textbf{2x laptop b} & Model & Dell G15 5521 \\
    & OS       & Windows 11 x64 \\
    & GPU  & NVIDIA GeForce RTX 3060 Mobile  \\
    & CPU  &  i7-12700H  \\ 
    \midrule
    \textbf{1x AP A} & Model & ASUS ROG Rapture GT-AXE11000 \\
    %&  Firmware    &  3.0.0.4.388\_22525 \\
    &  Standard               & 802.11ax         \\
    \textbf{2x AP~B/C} & Model & ASUS TUF-AX4200 \\
    &  Firmware    &  OpenWrt 23.05.5 \\
    &  Standard               & 802.11ax         \\
     \bottomrule
    \end{tabular}
    \label{tab:equipment}
\end{table}

%------------------------------------
%------------------------------------
%------------------------------------
%------------------------------------
\section{Experimental results}\label{sec:tests_results}

This section evaluates the impact of varying levels of OBSS contention on VR streaming performance using either CBR or NeSt-VR, across Scenarios~1 and~2.
Our evaluation considers both packet loss and VF-RTT. VF-RTT is a measure that captures network delay, defined as the elapsed time between the transmission of a video frame by the VR server and the reception of the client's acknowledgment confirming the frame arrival. As established in~\cite{casasnovas2025nest}, a 33~ms VF-RTT threshold is used as the upper bound for a satisfactory user experience at 90~fps streaming. In our testbed, under baseline, OBSS contention-free conditions, VF-RTTs averaged 5--6~ms.
Other latency components are not reported here, as their values remained consistent across all scenarios. Only game delay exhibited variation across scenarios, showing a similar trend to network delay due to its dependence on timely uplink data reception; nevertheless, it had a substantially smaller magnitude and minimal impact on overall latency.

\subsection{Coexistence with 80~MHz OBSSs (Scenario~1)}
Let us first consider Scenario~1, where one or two OBSSs operate on the same 80~MHz channel as the VR~BSS, sharing the same primary channel (52). This configuration leads to full channel overlap and direct contention for the medium.

\subsubsection{Full channel overlap with one contender}
 
A single OBSS contends with the VR~BSS on the same 80~MHz channel. Since both share the primary channel, their transmissions utilize the full 80~MHz bandwidth, enabling efficient and short data transmissions.

\begin{figure}[t]
    \centering    \includegraphics[width=.85\linewidth]{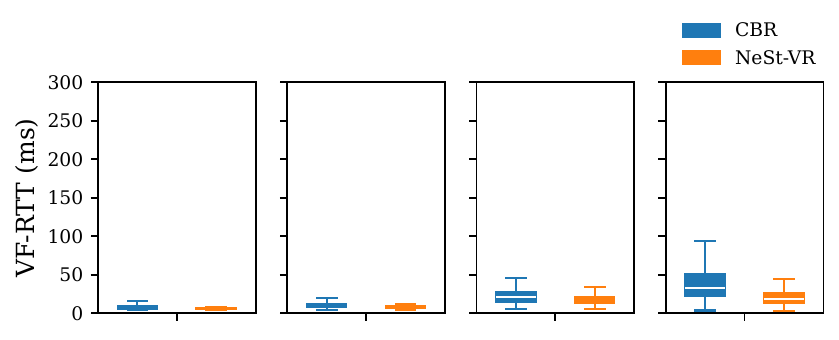}
    \includegraphics[clip, width=.85\linewidth, trim={0 0 0 0.65cm},]{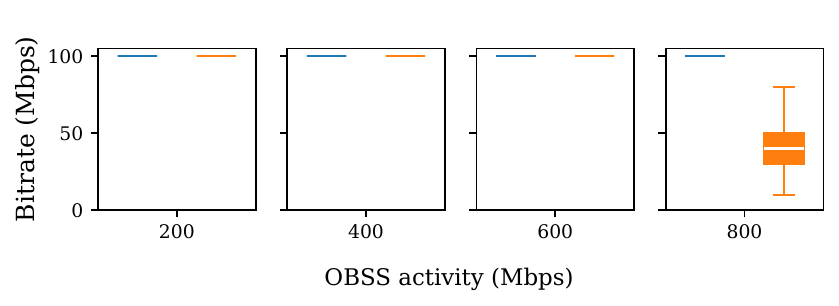}
    \caption{VF-RTT and target bitrate under Scenario~1, considering a single 80~MHz OBSS.
    }
    \label{fig:full_1}
\end{figure}

As illustrated in Fig.~\ref{fig:full_1}, at 200, 400, and 600~Mbps, both CBR and NeSt-VR maintain optimal performance: VF-RTTs remain consistently below 33~ms with no packet loss. Consequently, NeSt-VR maintains its target bitrate at 100~Mbps.
At 800~Mbps, packet loss also remains negligible; however, CBR experiences high VF-RTTs. In contrast, NeSt-VR mitigates user-perceivable delays by adaptively adjusting its bitrate to 35.1~Mbps~$\pm$~11.3 in 10~Mbps steps, effectively keeping latency within acceptable bounds.

\subsubsection{Full channel overlap with two contenders}\label{sec:full_2}
  
A second OBSS on the same channel is introduced, increasing contention. While all BSSs still use the full 80~MHz bandwidth, individual medium access opportunities decrease.

\begin{figure}[t]
    \centering
\includegraphics[width=.85\linewidth]{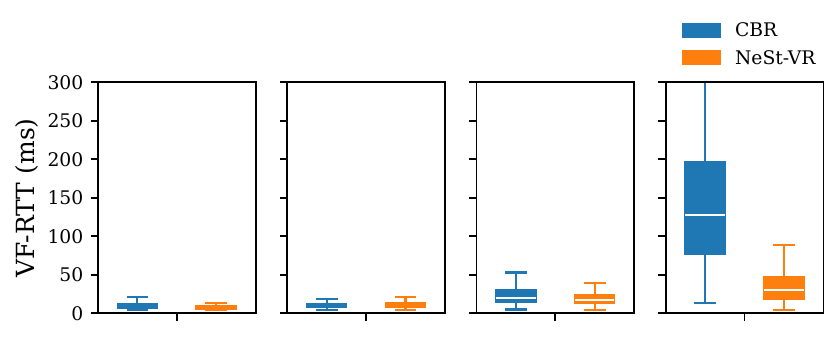}
    \includegraphics[clip, width=.85\linewidth, trim={0 0 0 0.65cm},]{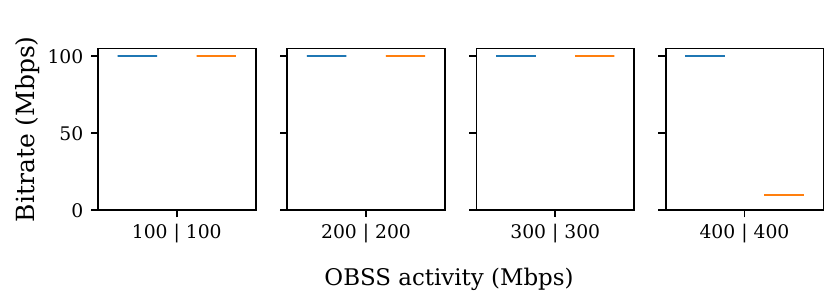}
    \caption{VF-RTT and target bitrate under Scenario~1, considering two 80~MHz OBSSs. $\text{OBSS A}~|~\text{OBSS B}$.
    }
    \label{fig:full_2}
\end{figure}

As depicted in Fig.~\ref{fig:full_2}, up to 600~Mbps aggregate (300~Mbps per OBSS), both schemes maintain VF-RTTs below 33~ms with no packet loss, thereby, requiring no bitrate adaptation using NeSt-VR.  
At 800~Mbps aggregate (400~Mbps per OBSS), CBR performance degrades significantly: VF-RTTs exceed hundreds of milliseconds (149.41~ms on average), and packet loss rises to 7.71\%, severely affecting the user experience. In contrast, NeSt-VR mitigates this degradation by reducing the streaming bitrate to the minimum (10~Mbps), achieving more acceptable VF-RTTs (36.8~ms on average) and minimal packet loss (0.16\%), although the network is clearly operating at its limit under these conditions. 

\subsubsection{Takeaways}\label{sec:full_discussion}

Compared to the single-contender case at 800~Mbps, the presence of two contenders leads to significantly higher contention and worse performance, despite the same total OBSS load. This confirms the intuitive expectation that the number of contending BSSs---not just their aggregate traffic---critically impacts performance due to increased airtime contention and collision probability.

\vspace{0.15cm}
%\hfill\\
\noindent\textit{\textbf{Takeaway 1:}}  
\textit{When considering OBSS contenders operating on the same 80~MHz channel as the VR~BSS, increasing their number intensifies contention and degrades VR streaming performance despite the same aggregate OBSS load.}

\noindent\textit{\textbf{Takeaway 2:}}  
\textit{NeSt-VR outperforms CBR under high OBSS contention. In Scenario~1, NeSt-VR maintains acceptable performance up to 800~Mbps aggregate OBSS load, while CBR only supports up to 600~Mbps, highlighting the value of adaptive bitrate control compared to the fixed-rate approach.}

%-------------------------------------
%-------------------------------------
\subsection{Coexistence with 40~MHz OBSSs (Scenario~2)}

Let us now consider Scenario~2, where one or two OBSS contenders operate on adjacent 40~MHz channels that partly overlap with the VR~BSS's 80~MHz channel. OBSS~A operates on the primary-sided 40~MHz portion, and OBSS~B on the secondary-sided portion. Accordingly, scenarios include partial channel overlap when a single contender is active and full channel overlap when both are active.

\subsubsection{Partial channel overlap with a primary-sided contender}
\label{sec:part_prim}

OBSS~A is the only active contender, operating on the primary-sided 40~MHz portion of the VR~BSS's 80~MHz channel. Therefore, the VR~BSS can detect OBSS~A's transmissions, deferring its activity accordingly. However, once the VR~BSS accesses the channel, it uses the full 80~MHz bandwidth.  

\begin{figure}[t]
    \centering
\includegraphics[width=.85\linewidth]{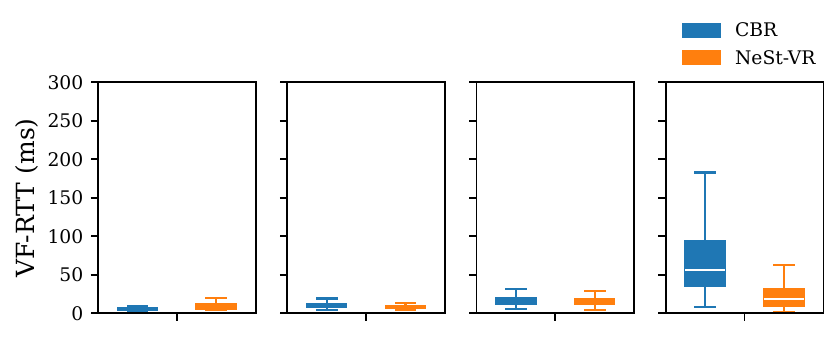}
    \includegraphics[clip, width=.85\linewidth, trim={0 0 0 0.65cm},]{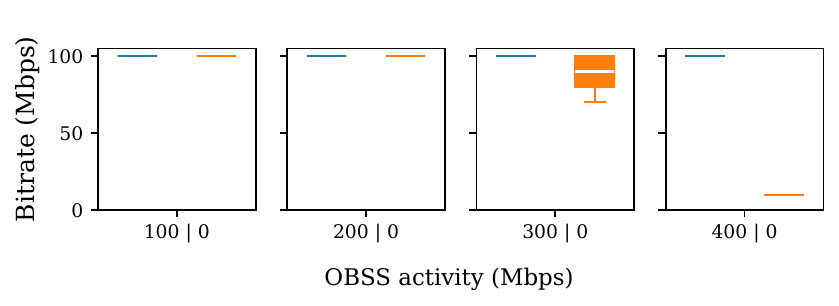}
    \caption{VF-RTT and target bitrate under Scenario~2, considering a single 40~MHz OBSS (OBSS~A).
    }
    \label{fig:partial_A}
\end{figure}

As shown in Fig.~\ref{fig:partial_A}, at 100 and 200~Mbps, both CBR and NeSt-VR maintain low VF-RTTs with no packet loss.  
At 300~Mbps, while packet loss remains negligible, CBR exhibits occasional VF-RTT spikes. NeSt-VR responds by slightly reducing its streaming bitrate to 90.97~Mbps~$\pm$~9.44, effectively maintaining latency below the 33~ms target.  
At 400~Mbps, CBR performance degrades significantly, with VF-RTTs exceeding hundreds of milliseconds (72.80~ms on average) and 0.49\% packet loss. In contrast, NeSt-VR reduces the streaming bitrate to the minimum (10~Mbps), effectively keeping latency within acceptable bounds (22.90~ms on average) and avoiding packet loss.  
Notably, OBSS~A achieves near-offered load throughput up to 300~Mbps regardless of whether CBR or NeSt-VR is used. At 400~Mbps, however, throughput drops to 256.2~Mbps under CBR, likely due to increased contention and resulting transmission inefficiencies. In contrast, under NeSt-VR, OBSS~A achieves 347.8~Mbps, since the reduced VR streaming bitrate lowers airtime demand, easing contention. Notably, even in isolation, OBSS~A does not fully reach 400~Mbps; it experiences around 10\% packet loss, likely due to hardware or software limitations.

%-------------------------------------

\subsubsection{Partial channel overlap with a secondary-sided contender} \label{sec:part_sec}

OBSS~B is the only active contender, operating on the secondary-sided 40~MHz portion of the VR~BSS's 80~MHz channel. As in the primary-sided case, the VR~BSS uses the full 80~MHz when OBSS~B is inactive. Conversely, when OBSS~B is active, the VR~BSS can transmit only over the primary-sided 40~MHz portion. This enables both BSSs to operate simultaneously on separate subchannels, increasing opportunities for concurrent transmissions.

As illustrated in Fig.~\ref{fig:partial_B}, the general trend is similar to the primary-sided case but with more pronounced degradation at higher offered loads. In particular, at 300~Mbps, CBR exhibits frequent and severe VF-RTT spikes, while NeSt-VR lowers its streaming bitrate more aggressively to 68.71~Mbps~$\pm$~6.19 to maintain latency.  
At 400~Mbps, the gap widens further: CBR suffers sustained VF-RTTs above several hundred milliseconds (115.05~ms on average) and 4.30\% packet loss. NeSt-VR again reduces its bitrate to the minimum (10~Mbps), avoiding packet loss and mitigating excessive delays (50.74~ms on average), though average VF-RTTs remain still above the 33~ms threshold, indicating persistent contention effects.  
Notably, OBSS~B achieves near-offered load throughput up to 200~Mbps. At 300~Mbps, throughput is slightly lower under CBR (275.6~Mbps) than NeSt-VR (285.9~Mbps). At 400~Mbps, throughput drops significantly under CBR to 236.1~Mbps, and even under NeSt-VR it remains below the offered load, reaching only 289.3~Mbps. These results confirm NeSt-VR's effectiveness in reducing contention, enabling the contender to achieve higher throughput while sustaining VR streaming performance. Interestingly, OBSS throughput (as VR performance) is consistently worse than in the primary-sided case.

Surprisingly, contention on the secondary-sided 40~MHz portion proves more detrimental to both the VR~BSS and the OBSS. This may result from ACI, potentially exacerbated in our setup, as OBSS~B AP is placed in close proximity to the VR~BSS AP. Energy spillover from nearby frequencies can cause devices to detect the channel as busy more frequently or experience degraded signal quality. This, in turn, may increase contention and/or reduce MCS rates, ultimately lowering throughput. Furthermore, OBSS~B's activity on the secondary-sided 40~MHz channel can force the VR~BSS to limit its channel to 20~MHz or defer access entirely. If the VR~AP is already transmitting, these overlapping signals may even corrupt ongoing VR~BSS transmissions.
Additionally, transmitting fewer but more efficient frames---whether by shortening their duration or by packing more data per transmission---over the full 80~MHz channel (as occurs with primary-sided contention) may be more effective than more frequent, narrower 40~MHz transmissions (as occurs during secondary-sided contention), especially if the OBSS contender is still granted sufficient access opportunities.

% ---------------------------------------
% ---------------------------------------
\subsubsection{Full channel overlap with two contenders}
 
Both OBSS~A and OBSS~B are active, each operating on a distinct 40~MHz portion of the VR~BSS's 80~MHz band, thereby jointly overlapping the full channel. In principle, OBSS~A and OBSS~B can be active simultaneously on their respective 40~MHz channels; conversely, the VR~BSS can use 80~MHz when neither OBSS is active, drop to 40~MHz (on the primary-sided portion) when only OBSS~B is active, and defer entirely whenever OBSS~A is active. It is considered both symmetric traffic loads (100, 200, or 300~Mbps each) and asymmetric loads (one at 100~Mbps, the other at 300~Mbps).

\begin{figure}[t]
    \centering
\includegraphics[width=.85\linewidth]{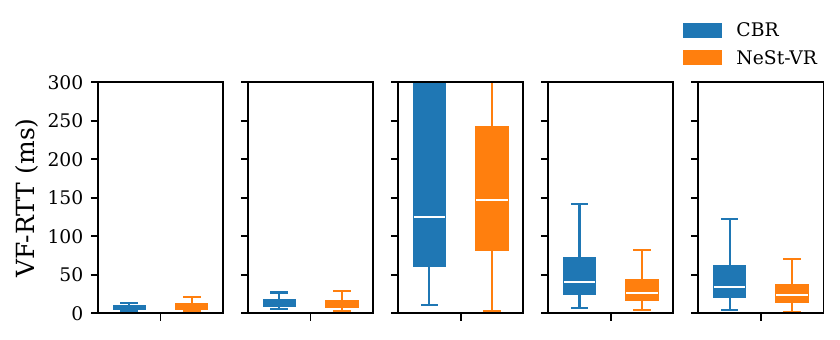}
    \includegraphics[clip, width=.85\linewidth, trim={0 0 0 0.65cm},]{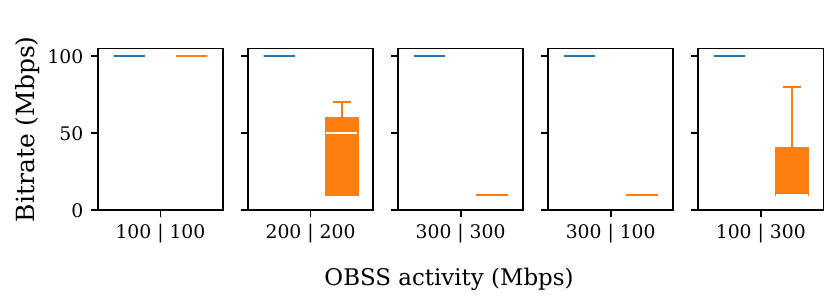}
    \caption{VF-RTT and target bitrate under Scenario~2, considering two 40~MHz OBSSs.
    }
    \label{fig:partial_2}
\end{figure}

As depicted in Fig.~\ref{fig:partial_2}, up to 400~Mbps aggregate load (200~Mbps per OBSS), VF-RTTs remain within acceptable bounds for both CBR and NeSt-VR, with no packet loss. Nevertheless, CBR shows occasional disruptive VF-RTT spikes, whereas NeSt-VR reduces its bitrate moderately to 42.26~Mbps~$\pm$~23.90 to avoid them.  
At 600~Mbps aggregate (300~Mbps per OBSS), CBR streaming degrades severely: VF-RTTs exceed hundreds of milliseconds (379.43~ms on average), and packet losses rise to 18.32\%, leading to video frame drops and even disconnections. In contrast, NeSt-VR reduces bitrate to the minimum (10~Mbps), significantly lowering packet loss (down to 0.43\%) and avoiding disconnections. Nevertheless, VF-RTTs remain well above acceptable thresholds (179.70~ms on average), as bitrate adaptation alone cannot fully mitigate degradation under such heavy contention.

\begin{figure}[t]
    \centering
    \includegraphics[width=.85\linewidth]{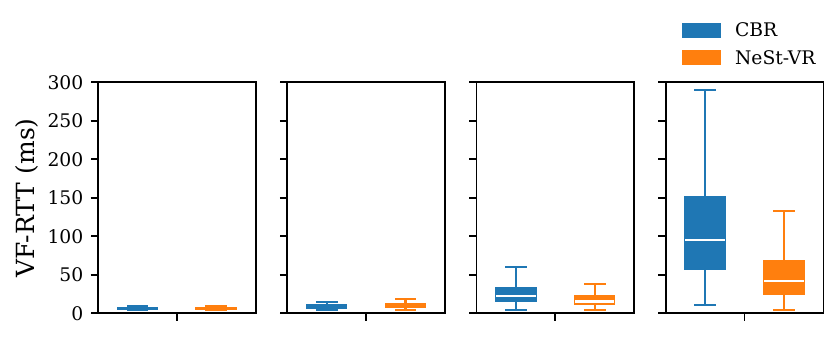}
     \includegraphics[clip, width=.85\linewidth, trim={0 0 0 0.65cm},]{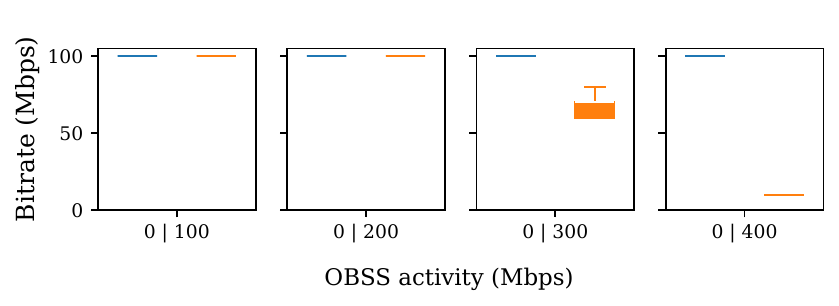}
    \caption{VF-RTT and target bitrate under Scenario~2, considering a single 40~MHz OBSS (OBSS~B).
    }
    \label{fig:partial_B}
\end{figure}

Interestingly, at 400~Mbps aggregate load, this case with both OBSSs outperforms the single-contender cases from Sections~\ref{sec:part_prim} and~\ref{sec:part_sec} (illustrated in Figs.~\ref{fig:partial_A} and~\ref{fig:partial_B}, respectively), suggesting that partial channel overlap with a single high-load 40~MHz contender can be more detrimental than full overlap with two moderate 40~MHz contenders occupying adjacent segments.  
Comparing these results to the full channel overlap with two 80~MHz OBSSs from Section~\ref{sec:full_2}, particularly at 600~Mbps aggregate, it appears that two 40~MHz contenders each on a distinct portion are more disruptive than two 80~MHz contenders. This suggests that partial overlaps can lead to more inefficient channel sharing, and that fully overlapping contenders may coordinate better through carrier sensing.
This observation aligns with~\cite{deek2011impact}, which found that fully overlapping contenders degrade performance less than partly overlapping ones, because they can sustain higher transmission rates while sharing the channel more fairly.  

Considering the asymmetric loads case, Fig.~\ref{fig:partial_2} shows that under CBR, VF-RTTs exceed the acceptable threshold, with minor but noticeable packet losses in both cases: 0.78\% when the heavier load is on OBSS~A, and 0.57\% when it is on OBSS~B.  
NeSt-VR adapts accordingly, adjusting the bitrate to 12.90~Mbps~$\pm$~6.93 and 25.81~Mbps~$\pm$~21.87, respectively. This eliminates packet loss and keeps VF-RTTs consistently below 33~ms once adaptation completes, again demonstrating the algorithm's effectiveness.

Interestingly, at the same aggregate load, this scenario with both OBSSs still outperforms the single-contender cases from Sections~\ref{sec:part_prim} and~\ref{sec:part_sec}.  
Based on those single-contender results, one might expect worse performance when the higher load is on the secondary-sided portion (OBSS~B) than on the primary-sided portion (OBSS~A). However, under CBR both asymmetric configurations showed similar levels of degradation, with even slightly better results when the heavier load was on OBSS~B. Indeed, NeSt-VR responded with more aggressive bitrate adjustments when the heavier load was on OBSS~A.  
This suggests that their individual impact is mitigated, likely because, in our setup, the OBSS APs are placed close together, leading to ACI that causes both OBSSs to contend not only with the VR~BSS but also with each other.

\subsubsection{Takeaways} \label{sec:part_discussion}

Overall, for the same aggregate load (i.e., 400~Mbps), asymmetric configurations resulted in worse VR~BSS performance than symmetric ones, regardless of which OBSS carried the higher load. This suggests that mutual OBSS contention indirectly leaves more opportunities for VR~BSS transmissions and, in the symmetric cases, enables more balanced channel sharing.

\vspace{0.15cm}
\noindent\textit{\textbf{Takeaway 3:}}  
\textit{An OBSS operating on the secondary-sided 40~MHz portion of the channel degrades VR~BSS streaming performance more than one on the primary-sided portion.}

\noindent\textit{\textbf{Takeaway 4:}}  
\textit{Full channel overlap with two moderate-load OBSS contenders, each on adjacent 40~MHz portions, can be less detrimental to VR streaming than partial overlap with a single high-load contender, for the same aggregate OBSS load.}

\noindent\textit{\textbf{Takeaway 5:}}  
\textit{Full channel overlap with two 40~MHz contenders on distinct adjacent portions is more disruptive to VR streaming than overlap with two 80~MHz contenders at the same aggregate OBSS load.}

\noindent\textit{\textbf{Takeaway 6:}}  
\textit{When two OBSS contenders occupy adjacent 40~MHz portions, symmetric OBSS loads are less detrimental to VR streaming than asymmetric loads for the same aggregate, as symmetric contention enables more balanced channel sharing. As Takeaway~4 this reinforces that distributing load evenly helps mitigate performance degradation.}

\noindent\textit{\textbf{Takeaway 7:}}  
\textit{NeSt-VR again outperforms CBR under high OBSS contention. In Scenario~2, NeSt-VR maintains acceptable performance up to 400~Mbps aggregate OBSS load, while CBR does so only when there are two OBSSs with symmetric activity (i.e., 200~Mbps per OBSS).}

%------------------------------------
%------------------------------------
%------------------------------------
%------------------------------------

\section{Conclusions}\label{sec:conclusions}

In this work, we experimentally analyzed the impact of OBSS contention on VR streaming over Wi-Fi~6 in two scenarios: one where all BSSs share the same 80~MHz channel, and another where contenders occupy distinct 40~MHz portions of the VR BSS's 80~MHz channel. 
Future research should investigate emerging Wi-Fi features that could further mitigate OBSS contention in dense scenarios, specifically Multi-Link Operation (MLO) (see~\cite{carrascosa2024performance}), which enables simultaneous transmissions across multiple bands or channels, and Multi-AP Coordination (MAPC) (see~\cite{nunez2025enabling}), which may facilitate coordinated contention and interference management.

\section{Acknowledgments}

This work is partially supported by Wi-XR PID2021-123995NB-I00 and TRUE-Wi-Fi PID2024-155470NB-I00 (MCIU/AEI/FEDER,UE), by MCIN/AEI under the Maria de Maeztu Units of Excellence Programme (CEX2021-001195-M), and by AGAUR ICREA Academia 000777.

%------------------------------------
%------------------------------------

\bibliographystyle{elsarticle-num} 
\bibliography{References}

\end{document}